\documentclass[12pt]{article}

         \usepackage{amsfonts}
         \usepackage{amssymb}

\def\be{\begin{eqnarray}}
\def\ee{\end{eqnarray}}
\def\bee{\begin{eqnarray*}}
\def\eee{\end{eqnarray*}}

\newtheorem{thm}{Theorem}

\newtheorem{lemma}[thm]{Lemma}

\newtheorem{defn}[thm]{Definition}

         \def\QED{{\bf QED}}

\def\bra{\langle}
\def\ket{\rangle}

\def\Tr{{\rm Tr}}

           \title{
The capacity of the quantum depolarizing channel}
\author{Christopher King
\\ Department of Mathematics
\\ Northeastern University
\\ Boston MA 02115
\\ {\normalsize and}
\\ Communications Network
Research Institute \\ Dublin Institute of Technology \\
  Rathmines, Dublin 6 \\
Ireland  \\
{\normalsize king@neu.edu}
}
\begin{document}

\maketitle

\begin{abstract}
The information carrying capacity of the $d$-dimensional
depolarizing channel is computed. It is shown that this capacity can be
achieved by encoding messages as products of pure states
belonging to an orthonormal basis of the state space, and using
measurements which are products of projections onto this same
orthonormal basis. In other words, neither entangled signal states
nor entangled measurements give any advantage for information capacity.
The result follows from an additivity theorem for the product
channel $\Delta \otimes \Psi$, where $\Delta$ is the depolarizing
channel and $\Psi$ is a completely arbitrary channel.
We establish the Amosov-Holevo-Werner $p$-norm conjecture
for this product channel for all $p \geq 1$, and deduce from this
the additivity of the minimal entropy and of the Holevo
quantity $\chi^{*}$.
\end{abstract}

\pagebreak



\section{Background and statement of results}

\subsection{Introduction}
This paper computes the capacity of the $d$-dimensional quantum depolarizing
channel for transmission of
classical information. The result confirms a longstanding conjecture,
namely that the best rate of information transfer can be achieved without
any entanglement across multiple uses of the channel.
It is sufficient to choose an orthonormal basis
for the state space, and then use this basis both to encode messages as
product states at the input side and to perform measurements at the
output side which project onto this same basis.
In this sense the depolarizing channel can be treated as a classical channel.

The Holevo-Schumacher-Westmoreland theorem allows the capacity to be
expressed in terms of the Holevo quantity $\chi^{*}$. If the Holevo
quantity is additive, then this expression implies that the capacity
is actually equal to $\chi^{*}$. In this paper we prove
several additivity properties for the depolarizing channel,
including the additivity of $\chi^{*}$. 
One important mathematical tool used in the proof is the Lieb-Thirring
inequality, which provides a bound for the non-commutative $p$-norm of 
a product of positive matrices.
These notions, as well as the definition of channel capacity
and its relation to the Holevo quantity, are described in the following
subsections.

\subsection{The depolarizing channel}
The depolarizing channel is a particularly simple model for noise in
quantum systems \cite{NC}, and has been studied in a variety of contexts
\cite{BDS, BSST, BFMP}. In $d$ dimensions the model
is implemented by a completely positive trace-preserving map
$\Delta_{\lambda}$, depending on one real parameter $\lambda$, which
maps a state $\rho$ on ${\bf C}^d$ into a linear combination
of itself and the $d \times d$ identity matrix $I$:
\be\label{def:Delta}
\Delta_{\lambda}(\rho) =  \lambda \, \rho + {1 - \lambda \over d} \, I
\ee
The condition
of complete positivity requires that $\lambda$ satisfy the bounds
\be\label{range-dep}
-{1 \over d^2 - 1} \leq \lambda \leq 1
\ee

The channel $\Delta_{\lambda}$ maps a pure input state
to a mixed output state. Because the channel is highly symmetric,
all such output states are unitarily equivalent, and have eigenvalues
$\lambda + (1 - \lambda)/d$ (with multiplicity 1) and
$(1 - \lambda)/d$  (with multiplicity $d-1$).

\subsection{Measures of noisiness}
We will use three measures of noisiness for quantum channels,
namely
the minimal output entropy $S_{\rm min}$, the maximal output $p$-norm
$\nu_{p}$, and the Holevo quantity $\chi^{*}$.
First recall that
a state $\rho$ is a positive operator with trace equal to $1$, and its
von Neumann entropy is defined by
\be
S(\rho) = - \Tr \rho \log \rho
\ee
The minimal output entropy of the channel $\Psi$  is
defined as
\be\label{def:S-min}
S_{\rm min}(\Psi) = \inf_{\rho} S\big(\Psi(\rho)\big)
\ee
Recall also that the $p$-norm of a positive matrix $A$ is defined
for $p \geq 1$  by
\be\label{def:p-norm}
|| A ||_{p} = \Big( \Tr A^p \Big)^{1/p}
\ee
Amosov, Holevo and Werner \cite{AHW} introduced the notion of the
maximal $p$-norm of a channel as a way to characterize its noisiness.
This quantity is defined as
\be\label{def:nu}
\nu_{p}(\Psi) = \sup_{\rho} || \Psi(\rho) ||_{p}
\ee
Since the entropy of a state is the negative of the derivative of the
$p$-norm at $p=1$, it follows that for any channel $\Psi$,
\be
{d \over d p} \nu_{p}(\Psi) \Big{|}_{p=1} = - S_{\rm min}(\Psi)
\ee

The third measure of noisiness, the Holevo quantity, is
closely related to the information-carrying capacity of the channel, as will
be explained in the next section.
We  will use the symbol $\cal E$ to
denote an ensemble
of input states for the channel, that is a collection of states $\rho_i$
together with
a probability distribution $\pi_i$. The Holevo quantity is
\be\label{def:chi}
\chi^{*}(\Psi) = \sup_{\cal E} \Big[ S\big(\Psi(\rho)\big)
- \sum_{i} \pi_{i} S\big(\Psi(\rho_i)\big) \Big]
\ee
where $\rho = \sum \pi_i \rho_i$ is the average input state of the
ensemble.

These three measures can be easily computed for the depolarizing channel.
The values are
\be\label{s-min}
S_{\rm min}(\Delta_\lambda) = - \Big( \lambda + {1 - \lambda \over d} \Big)
\log \Big(
\lambda + {1 - \lambda \over d} \Big) - (d-1) ({1 - \lambda \over d}) \log
\Big(
{1 - \lambda \over d} \Big)
\ee
\be\label{nu4dep}
\nu_{p}(\Delta_{\lambda}) = \bigg[ \Big( \lambda + {1 - \lambda \over d}
\Big)^p +
(d-1) \Big({1 -
\lambda \over d} \Big)^p
\bigg]^{1/p}
\ee
\be\label{chi4dep}
\chi^{*}(\Delta_\lambda) = \log d - S_{\rm min}(\Delta_\lambda)
\ee
The value (\ref{chi4dep}) is achieved by choosing an ensemble
$\cal E$ consisting of  pure states belonging to an orthonormal basis
(it does'nt matter which one),
and choosing the uniform distribution $\pi_i = 1/d$. Since the
average input state for this ensemble is $(1/d) \, I$, the terms on
the right side of (\ref{def:chi}) are separately maximized for this choice of
ensemble, and this leads to the result (\ref{chi4dep}).

\subsection{Additivity conjectures}
It is conjectured that $S_{\rm min}$ and $\chi^{*}$ are both additive for
product channels. This would mean that for any channels $\Psi_1$ and $\Psi_2$,
\be\label{conj1}
S_{\rm min}(\Psi_1 \otimes \Psi_2) = S_{\rm min}(\Psi_1) +  S_{\rm
min}(\Psi_2)
\ee
and
\be\label{conj2}
\chi^{*}(\Psi_1 \otimes \Psi_2) = \chi^{*}(\Psi_1) + \chi^{*}(\Psi_2)
\ee
Equivalently, the conjectures would imply that for product channels
both of these
measures of noisiness are achieved  with product input states.
As we explain in the next section,
the Holevo quantity $\chi^{*}$ is related to channel capacity,
and in fact (\ref{conj2}) would imply that the channel capacity
of an arbitrary channel $\Psi$ is
precisely $\chi^{*}(\Psi)$.
These conjectures have been established for some special
classes of channels, including all unital qubit channels \cite{Ki2}
and all entanglement-breaking channels \cite{Shor}. However
a general proof has remained elusive.

It was further conjectured in \cite{AHW} that the quantity $\nu_{p}$ should be
multiplicative for product channels for all $p \geq 1$,
reflecting the idea that the $p$-norm of a
product channel would be maximized with product states. This
conjecture would imply  (\ref{conj1}), since the entropy can be obtained
from the derivative  of the $p$-norm at $p=1$.
Again this conjecture has been established for
some special classes of channels, in particular for unital qubit channels
\cite{Ki2}.
However it is now known that the conjecture does not hold in general --
this was demonstrated recently
by the discovery of a family of counterexamples for values $p \geq 5$ \cite{Werner}.
Nevertheless, in this paper we will prove that the AHW conjecture is true for any
product channel of the form $\Delta_{\lambda} \otimes \Psi$
where $\Psi$ is arbitrary, and we will show how this property implies the
additivity of
$S_{\rm min}$ and of $\chi^{*}$ for such a product channel.

\subsection{Channel capacity}
In order to relate these additivity results to the channel capacity problem,
we recall first the definition of the capacity for a general channel
$\Psi$.  Again we denote by $\cal E$ an ensemble of input states,
and we also denote by $\cal M$ a measurement, or POVM, at the output side of
the channel. Recall that a POVM
is a collection of positive operators
$\{ E_{j} \}$ which sum to the identity matrix. When a state $\rho$
is measured using a POVM $\{ E_{j} \}$, the outcome $j$ is obtained
with probability $\Tr (\rho E_j)$. This notion generalizes the familiar
von Neumann measurement, which is the special case where the operators
$E_j$ are  orthogonal projections.

The ensemble $\cal E$, the POVM $\cal M$ and the channel
$\Psi$ together define a
classical noisy channel, whose transition matrix is
\be
p_{i j} = \Tr \Big[ \Psi(\rho_i) E_j \Big]
\ee
If we write $X$ for a random input signal with distribution $\pi_i =
P(X~=~i)$,
then the output signal $Y$ from this classical channel has distribution
\be\label{def:Y}
P(Y~=~j) = \sum_{i} \pi_i p_{i j}
\ee
The Shannon capacity of a classical noisy channel measures the maximum rate
at which information can be reliably transmitted through the channel.
Shannon's formula computes this capacity as the maximum of the mutual
information $I(X, Y)$ between an input signal $X$ and its corresponding output
signal $Y$ given by (\ref{def:Y}), where the maximum is evaluated over all
choices of distribution $\{ \pi_i \}$ for $X$.
For the case of a quantum channel, we are
interested in the maximum rate that can be achieved using the optimal choices
of input states $\{ \rho_i \}$ and of output measurements $\{ E_j \}$.
Therefore we are led to define the Shannon capacity of the quantum channel
$\Psi$ as
\be\label{def:C-Shan}
C_{\rm Shan}(\Psi) = \sup_{{\cal E}, {\cal M}} \, I(X, Y)
\ee
where the distribution of the input signal $X$ is determined by the ensemble
$\cal E$, that is $P(X~=~i) = \pi_i$, and where the output signal $Y$ is
determined by (\ref{def:Y}).
This can be easily evaluated for the depolarizing channel. The mutual
imformation $I(X, Y)$ in (\ref{def:C-Shan}) is maximized by choosing an
ensemble consisting
of projections onto an orthonormal basis, and using the same
basis for the measurement. The result is
\be\label{eval}
C_{\rm Shan}(\Delta_{\lambda}) = \chi^{*}(\Delta_{\lambda})
\ee
where the capacity $\chi^{*}(\Delta_{\lambda})$ is given by
(\ref{chi4dep}).

If two copies of the channel $\Psi$ are available then it may be possible
to achieve a higher rate of transmission by sharing the signals across the two
channels. This possibility exists because
quantum channels have an additional resource which is not
available for classical channels, namely entangled states
which can be used to encode signals for the product channel. It is also
possible to make measurements at the output side using a POVM
which projects onto entangled states.
With these resources the best rate that can be achieved using two copies
of the channel is
\be\label{2 uses}
{1 \over 2} C_{\rm Shan}(\Psi \otimes \Psi)
\ee
It is known that in general (\ref{2 uses}) is larger than (\ref{def:C-Shan})
\cite{Hol2, BFS}.
This observation leads to the question of finding the asympototic capacity
which
would be achieved
by sharing the input signals across an unlimited number of copies of $\Psi$.
This ultimate capacity is given by
\be\label{def:c-ult}
C_{\rm ult}(\Psi) = \lim_{n \rightarrow \infty} {1 \over n}
 C_{\rm Shan}(\Psi^{\otimes n})
\ee
(a standard subadditivity argument shows the existence of this limit).

At the present time it is an open problem to determine $C_{\rm ult}(\Psi)$
for an arbitrary channel $\Psi$. However it can be expressed in terms of
the Holevo
quantity (\ref{def:chi}). Recall that the Holevo bound implies that
\be
C_{\rm Shan}(\Psi) \leq \chi^{*}(\Psi),
\ee
and hence that
\be\label{upper}
C_{\rm ult}(\Psi) \leq \lim_{n \rightarrow \infty} {1 \over n}
\chi^{*}(\Psi^{\otimes n})
\ee
Furthermore the Holevo-Schumacher-Westmoreland theorem \cite{Hol1, SW1}
shows that the rate
$\chi^{*}(\Psi)$ can be achieved with multiple copies of the channel,
by restricting to product states for the input signals, but allowing entangled
measurements at the outputs. Applying this theorem to
the product channel $\Psi^{\otimes n}$ implies that the rate $(1/n) \,
\chi^{*}(\Psi^{\otimes n})$
is achieved with input signals which may be entangled across $n$ uses of the
channel. Allowing $n$ to be arbitrarily large leads to the result
\be\label{equal}
C_{\rm ult}(\Psi) = \lim_{n \rightarrow \infty} {1 \over n}
\chi^{*}(\Psi^{\otimes n})
\ee

\subsection{Statement of results}
Our first result is the evaluation of (\ref{def:c-ult}) for the
depolarizing channel.

\medskip
\begin{thm}\label{thm1}
The capacity of the $d$-dimensional depolarizing channel is
\be
C_{\rm ult}(\Delta_\lambda) = \chi^{*}(\Delta_\lambda) =
C_{\rm Shan}(\Delta_\lambda) = \log d - S_{\rm min}(\Delta_\lambda)
\ee
where $S_{\rm min}(\Delta_\lambda)$ is evaluated in (\ref{s-min}).
\end{thm}

\medskip

The fact that $C_{\rm ult}(\Delta_\lambda) = C_{\rm Shan}(\Delta_\lambda)$
means that as far as the information-carrying properties of the depolarizing
channel are concerned, there is no advantage gained by using either entangled
input states or using entangled measurements. The optimal rate can be
achieved by
choosing an orthonormal basis (because $\Delta_{\lambda}$ is symmetric
it does'nt matter which one) and using this
basis to encode the signals and also to measure them. In this sense the
channel behaves
like a classical channel, and entanglement does not play any role in its
capacity.

The basic ingredient in the proof of Theorem \ref{thm1}
is the additivity of the Holevo quantity $\chi^{*}$ for the depolarizing
channel,
which we state in the next Theorem.

\medskip
\begin{thm}\label{thm2}
For any channel $\Psi$,
\be\label{eqn1thm2}
\chi^{*}(\Delta_\lambda \otimes \Psi) = \chi^{*}(\Delta_\lambda) +
\chi^{*}(\Psi)
\ee
\end{thm}

\medskip
Theorem \ref{thm1} follows easily from this, as we now demonstrate.
The result $C_{\rm ult}(\Delta_\lambda) = \chi^{*}(\Delta_\lambda)$ in
Theorem \ref{thm1} follows immediately from Theorem \ref{thm2}
by choosing $\Psi = \Delta_\lambda^{\otimes n}$ in (\ref{eqn1thm2}) and
applying (\ref{equal}). The second equality
$\chi^{*}(\Delta_\lambda) = C_{\rm Shan}(\Delta_\lambda)$ was
derived in (\ref{eval}).

\medskip

Finally we state the AHW conjecture for the depolarizing
channel, which underlies all the other results.
Since the derivative of $\nu_{p}(\Psi)$ at $p=1$ is equal to $- S_{\rm
min}(\Psi)$,
the additivity of $S_{\rm min}$ is a special case of the
AHW conjecture. We state both results  next in Theorem \ref{thm3}.

\medskip
\begin{thm}\label{thm3}
For any channel $\Psi$, and any $p \geq 1$,
\be\label{AHW}
\nu_{p}(\Delta_\lambda \otimes \Psi) = \nu_{p}(\Delta_\lambda) \,
\nu_{p}(\Psi)
\ee
and hence
\be
S_{\rm min}(\Delta_\lambda \otimes \Psi) = S_{\rm min}(\Delta_\lambda)
+ S_{\rm min}(\Psi)
\ee
\end{thm}

\medskip
Special cases of Theorem \ref{thm3} were previously established, namely
for integer values of $p$ in all dimensions $d$ \cite{AH}, and for  all $p
\geq 1$
in dimension $d=2$ \cite{Ki2}.

\subsection{Organization}
The paper is organised as follows. Section 2 outlines the proof of
Theorem \ref{thm3}, and states two key results which are used, namely
the convex decomposition of the depolarizing channel, and the bound
for the phase-damping channel. These results are then established in
Sections 3 and 4, and finally Theorem \ref{thm2} is proved in
Section 5. Section 3 also describes in detail the convex
decompositions for the two-dimensional  qubit depolarizing channel.
Section 6 contains some discussion of the nature of
the proof, and some ideas about further directions to pursue.

\section{Outline of the proof}
\subsection{Definition of phase-damping channel}
As discussed above, Theorem \ref{thm1} follows immediately from
Theorem \ref{thm2}, using the
HSW Theorem (\ref{equal}). Theorem \ref{thm2} itself is a slight extension of
Theorem \ref{thm3}, and will be proved in Section 5.
Most of the work in this paper goes into
the proof of
the AHW conjecture (\ref{AHW}) in Theorem \ref{thm3}.
The proof presented here
develops further the methods introduced in \cite{Ki2} where the
same result was established for unital qubit channels.
The basic idea is similar: we express the depolarizing channel
$\Delta_\lambda$ as a convex combination of simpler channels,
and then we prove a bound for these simpler channels which
implies the result (\ref{AHW}). In \cite{Ki2} the simpler channels
were unitarily equivalent to phase-damping channels, and we
use the same name for the channels here, which  are defined
as follows.

\medskip
\begin{defn}\label{def1}
Let ${\cal B} = \{ | \psi_i \ket \}$ be an orthonormal basis, and
let $E_i = | \psi_i \ket \bra \psi_i |$. The phase-damping
channel corresponding to $\cal B$ is the one-parameter family of maps
\be\label{def:ph-d}
\Phi_{\lambda}(\rho) = \lambda \rho + (1 - \lambda) \sum_{i=1}^d  E_i \rho E_i
\ee
where the parameter $\lambda$ satisfies the bounds
\be\label{range-ph}
-{1 \over d - 1} \leq \lambda \leq 1
\ee
\end{defn}

\medskip
The parameter range (\ref{range-ph}) is required by the condition of
complete positivity. If we write $\rho = (\rho_{ij})$
as a matrix in the basis
$| \psi_i \ket$ then the channel (\ref{def:ph-d}) acts by scaling the
off-diagonal entries and leaving unchanged the diagonal entries, that is
\be\label{diag-basis}
\Phi_{\lambda}(\rho)_{ij} = \cases{\rho_{ij} & if $i = j$ \cr
\lambda \rho_{ij} & if $i \neq j$ \cr}
\ee

We will express $\Delta_{\lambda}$ as a convex
combination of phase-damping
channels, all with the same parameter $\lambda$. Furthermore
these phase-damping channels will all share a common
property, which is expressed by the following
definition.

\medskip
\begin{defn}\label{def5}
We say that a vector ${\bf v} = (v_1, \dots, v_d)$ in ${\bf C}^d$
is {\bf uniform} if $|v_i| = |v_j|$ for all
$i,j = 1, \dots ,d$.
\end{defn}
\medskip

It will turn out that the phase-damping channels which
arise in the convex decomposition are constructed from
orthonormal bases ${\cal B} = \{ | \psi_i \ket \}$ where
all the vectors $| \psi_i \ket$ are uniform. Since each vector
$| \psi_i \ket$ is normalized, it follows that all its entries
have absolute value $1/\sqrt{d}$.
As a consequence,
if $D$ is any diagonal matrix, then for any uniform
state $| \psi \ket$ 
\be\label{diag-D}
\Tr \Big[ | \psi \ket \bra \psi | \, D \Big] =
\bra \psi | \, D \, | \psi \ket = {1 \over d} \, \Tr D
\ee

\medskip
\begin{defn}\label{def2}
Let $\Phi$ be the phase-damping channel corresponding to the orthonormal
basis $\cal B$.
We say that $\Phi$ is {\bf uniform}
if $| \psi_i \ket$  is uniform for every $| \psi_i \ket \in {\cal B}$.
\end{defn}
\medskip

\subsection{Three lemmas}
There are three steps in the proof of Theorem \ref{thm3}.
The goal is to find a bound for
$|| (\Delta_{\lambda} \otimes \Psi) (\tau_{12}) ||_{p}$ which will
lead to (\ref{AHW}),
where $\Psi$ is any other channel, and $\tau_{12}$ is any state.
The first step is a partial diagonalization of the state
$\tau_{12}$. This step uses the following invariance property
of the depolarizing channel.

\medskip
\begin{lemma}\label{lemma1}
Let $\tau_1 = {\Tr}_{2} (\tau_{12})$ denote the reduced
density matrix of $\tau_{12}$, and let $U$ be a unitary
matrix. Define $\tau_{12}' = (U \otimes I) \tau_{12}
(U^{*} \otimes I)$. Then for all $p \geq 1$
\be\label{l1}
|| (\Delta_{\lambda} \otimes \Psi) (\tau_{12}) ||_{p}
= || (\Delta_{\lambda} \otimes \Psi) (\tau_{12}') ||_{p}
\ee
\end{lemma}
\medskip
\noindent{\bf Proof}: the definition of $\Delta_{\lambda}$
in (\ref{def:Delta}) implies that the unitary matrix $U \otimes I$
can be pulled through the channel $\Delta_{\lambda} \otimes \Psi$,
and then the invariance of the $p$-norm implies (\ref{l1}). \QED

\medskip

The second step uses the following result which expresses
$\Delta_{\lambda}$ as a convex combination of phase-damping
channels. This result will be derived in Section 3.
\medskip
\begin{lemma}\label{lemma2}
For $n=1, \dots, 2d^{2}(d+1)$,
there are positive numbers $c_n$, unitary matrices
$U_n$ and uniform phase-damping channels $\Phi_{\lambda}^{(n)}$
such that for any state $\rho$
\be\label{conv1}
\Delta_{\lambda}(\rho) = \sum_{n=1}^{2d^{2}(d+1)} c_{n} \,
U_{n}^{*} \, {\Phi_{\lambda}}^{(n)}(\rho)  \, U_{n}
\ee
\end{lemma}

\medskip
The third step in the proof uses the following bound for the
phase-damping channels. This bound will be derived in Section 4.

\medskip

\begin{lemma}\label{lemma3}
Let $\Phi_{\lambda}$ be a phase-damping channel defined as in
(\ref{def:ph-d}),
with corresponding orthogonal projectors $E_i = | \psi_i \ket \bra \psi_i |$.
For an arbitrary bi-partite state $\rho_{12}$ define
\be\label{def:rho^i}
\rho_{2}^{(i)} = {\Tr}_{1} \bigg[ (E_i \otimes I)
\rho_{12} \bigg]
\ee
 where ${\Tr}_{1}$ is the trace over the first
factor.  Recall the factor
$\nu_{p}(\Delta_{\lambda})$ from (\ref{nu4dep}).
Then for all $p \geq 1$,
\be\label{bound1}
|| (\Phi_{\lambda} \otimes I)(\rho_{12}) ||_{p} \leq
d^{(1 - 1/p)} \, \, \nu_{p}(\Delta_{\lambda}) \,\,
\Big[ \sum_{i=1}^{d} \,
\Tr \big( \rho_{2}^{(i)} \big)^p \Big]^{1/p}
\ee

\end{lemma}

\subsection{Proof of Theorem \ref{thm3}}
We will now prove Theorem \ref{thm3} using these three lemmas.
Since the left side of (\ref{AHW}) is at least as big as the right
side, it is sufficient to prove that for any bipartite state $\tau_{12}$
\be\label{AHWbound}
|| (\Delta_{\lambda} \otimes \Psi)(\tau_{12}) ||_{p} \leq
\nu_{p}(\Delta_{\lambda}) \, \nu_{p}(\Psi)
\ee

The first step is to use Lemma 1 to partially diagonalize the state
$\tau_{12}$. Let $U$ be a unitary matrix which
diagonalizes the reduced density matrix $\tau_{1} = {\Tr}_{2}
(\tau_{12})$, and let $\tau_{12}' = (U \otimes I) \tau_{12}
(U^{*} \otimes I)$, so that $\tau_{1}' = U \tau_{1}
U^{*}$ is diagonal. By Lemma \ref{lemma1} we can replace $\tau_{12}$ by
$\tau_{12}'$ without changing the left side of (\ref{AHWbound}).
Therefore we will assume henceforth without loss of generality that
$\tau_{1}$ is diagonal.

\medskip

The second step is to apply the convex decomposition (\ref{conv1}) on
the left side of (\ref{AHWbound}):
\be\label{conv2}
(\Delta_{\lambda} \otimes \Psi) (\tau_{12}) =
\sum_{n=1}^{2d^{2}(d+1)} c_{n} \,
(U_{n}^{*} \otimes I) \, ({\Phi_{\lambda}}^{(n)} \otimes \Psi)(\tau_{12})
\, (U_{n} \otimes I)
\ee

\medskip
For the third step, notice that
by convexity of the $p$-norm it is sufficient to prove the
bound (\ref{AHWbound}) for each term
$({\Phi_{\lambda}}^{(n)} \otimes \Psi)(\tau_{12})$ appearing on the
right side of (\ref{conv2}), namely
\be\label{AHW2}
|| (\Phi_{\lambda}^{(n)} \otimes \Psi)(\tau_{12}) ||_{p} \leq
\nu_{p}(\Delta_{\lambda}) \, \nu_{p}(\Psi)
\ee
In order to derive (\ref{AHW2}), we
apply (\ref{bound1}) with
\be
\rho_{12} = (I \otimes \Psi) (\tau_{12}), \quad
\rho_{2}^{(i)} = \Psi(\tau_{2}^{(i)}) =
\Psi \bigg( {\Tr}_{1} \Big[ (E_i \otimes I) \tau_{12} \Big] \bigg)
\ee
Therefore (\ref{bound1}) gives
\be\label{AHW3}
|| (\Phi_{\lambda}^{(n)} \otimes \Psi)(\tau_{12}) ||_{p} \leq
d^{(1 - 1/p)} \, \, \nu_{p}(\Delta_{\lambda}) \,\,
\Big[ \sum_{i=1}^{d} \,
\Tr \big( \Psi(\tau_{2}^{(i)}) \big)^p \Big]^{1/p}
\ee
Now the definition of the $p$-norm $\nu_{p}(\Psi)$ implies that for each $i$,
\be
\bigg[ \Tr \big( \Psi(\tau_{2}^{(i)}) \big)^p \bigg]^{1/p}  \leq
\nu_{p}(\Psi) \,\,
 \Tr (\tau_{2}^{(i)})
\ee
From the definition of $\tau_{2}^{(i)}$ it follows that
\be
\Tr (\tau_{2}^{(i)}) = \Tr (E_{i} \tau_{1})
\ee
Furthermore in the first step we chose the state $\tau_1$ to be diagonal, and
from Lemma \ref{lemma2} the phase-damping channels appearing
on the right side of (\ref{conv2}) are all uniform (recall
Definition \ref{def2}). Hence from (\ref{diag-D}) we get
\be
\Tr (E_{i} \tau_{1}) = {1 \over d} \, \Tr \tau_{1}
= {1 \over d}
\ee
Inserting into (\ref{AHW3}) gives
\be
|| (\Phi_{\lambda}^{(n)} \otimes \Psi)(\tau_{12}) ||_{p} & \leq &
d^{(1 - 1/p)} \, \, \nu_{p}(\Delta_{\lambda}) \, \nu_{p}(\Psi)
\Big[ d \, \bigg({1 \over d}\bigg)^p \Big]^{1/p} \\
& = & \nu_{p}(\Delta_{\lambda}) \, \nu_{p}(\Psi)
\ee
which completes the proof.
\QED

\section{The convex decomposition}

\subsection{Proof of Lemma \ref{lemma3}}
The derivation proceeds in two stages. First
we define a new channel $\Omega_{\lambda}$
which appears in an intermediate role:
\be\label{def:Omega}
\Omega_{\lambda}(\rho) = \Delta_{\lambda}(\rho) + {1 - \lambda \over d}
\bigg[ \rho - {\rm diag}(\rho) \bigg]
\ee
where ${\rm diag}(\rho)$ is the diagonal part of the matrix $\rho$.
It is not hard to see that $\Omega_{\lambda}$ is completely positive
and trace-preserving for all $\lambda$ in the range (\ref{range-ph}).
For example it can be re-written as
\be
\Omega_{\lambda}(\rho) =
 \Big(\lambda + {1-\lambda \over d}\Big) \rho
+ {(d-1)(1-\lambda) \over d} {1 \over d-1} \Big[I -
{\rm diag}(\rho) \Big],
\ee
and the map $\Big[I - {\rm diag}(\rho) \Big]/(d-1)$ is easily seen to
be completely positive and trace-preserving.
Next let $G$ be the diagonal unitary matrix
with entries
\be\label{def:G}
G_{kk} & = &  \exp \Big({2 \pi i k \over d}\Big), \quad 1 \leq k \leq d \\
G_{kl} & = & 0, \quad\quad k \neq l
\ee

\medskip

\begin{lemma}\label{lemma5}
For any matrix $\rho$,
\be\label{convex1}
\Delta_{\lambda}(\rho) = {\lambda d \over 1 + (d-1) \lambda} \,
\Omega_{\lambda}(\rho)
+ {1 - \lambda \over 1 + (d-1) \lambda}  \, {1 \over d} \, \sum_{k=1}^{d}
\big( G^{*} \big)^{k} \Omega_{\lambda}(\rho) G^{k}
\ee
\end{lemma}

\medskip
\noindent{\bf Proof:}
A straightforward  computation shows that for any matrix $\rho$
\be
{1 \over d} \, \sum_{k=1}^{d} \big( G^{*} \big)^{k} \, \rho \, G^{k}
= {\rm diag}(\rho)
\ee
Applying the definitions of $\Delta_{\lambda}$ and
$\Omega_{\lambda}$ from (\ref{def:Delta}) and
(\ref{def:Omega}), the result now follows easily.
\QED

\medskip

For the second stage in the derivation of (\ref{conv1})
we express the channel $\Omega_{\lambda}$ itself as a
convex combination of phase-damping channels.
To this end define the diagonal unitary matrix $H$ by
\be\label{def:H}
H_{kk} & = &  \exp \Big( {2 \pi i k^2  \over 2d^{2}} \Big), \quad 1 \leq k \leq d \\
H_{kl} & = & 0, \quad\quad k \neq l
\ee
and define the following pure state $| \theta \ket$:
\be\label{def:theta}
| \theta \ket = {1 \over \sqrt{d}} \, \pmatrix{1 \cr1 \cr
\vdots \cr
1 }
\ee
For each $k = 1, \dots, d$ and $a = 1, \dots, 2 d^2$
we define the pure state
\be\label{def:psi}
| \psi_{k,a} \ket = G^{k} H^{a} | \theta \ket
\ee
and the corresponding orthogonal projection
\be
E_{k,a} = | \psi_{k,a} \ket  \bra \psi_{k,a} |
\ee
For each fixed $a$, the states $\{ | \psi_{k,a} \ket \}$ form an orthonormal
basis. We denote by $\Phi_{\lambda}^{(a)}$ the corresponding family
of phase-damping channels, that is
\be\label{def:newPhi}
\Phi_{\lambda}^{(a)} (\rho) = \lambda \rho + (1 - \lambda)
\sum_{k=1}^d E_{k,a} \rho E_{k,a} ,
\quad \quad
a = 1, \dots , 2 d^2
\ee

\medskip
\begin{lemma}\label{lemma6}
\be\label{convex2}
\Omega_{\lambda} = {1 \over 2d^{2}} \, \sum_{a=1}^{2d^{2}} \Phi_{\lambda}^{(a)}
\ee
\end{lemma}

\medskip
\par\noindent
{\bf Proof:}
Using the definitions of $\Omega_{\lambda}$, $\Delta_{\lambda}$ and
$\Phi_{\lambda}^{(a)}$, it suffices to show that for any state $\rho$
\be
{1 \over 2d} \sum_{a=1}^{2d^{2}} \sum_{k=1}^d E_{k,a} \rho E_{k,a}
= I + \rho - {\rm diag}(\rho)
\ee
For each $x = 1, \dots, d$, let
$| x \ket$ be the unit vector with entry $1$ in position $x$, and
$0$ elsewhere. Then it suffices to show that for all $x,y$
\be\label{eqn1}
{1 \over 2d} \sum_{a=1}^{2d^{2}} \sum_{k=1}^d \, \bra x | E_{k,a} \rho E_{k,a} | y \ket
= \cases{1 & if $x=y$ \cr
\bra x |  \rho  | y \ket & if $x \neq y$
}
\ee
The $(a,k)^{\rm th}$ term on the left side of (\ref{eqn1}) can be written as
\be\label{(a,k)}
\bra x | E_{k,a} \rho E_{k,a} | y \ket
& = & \bra x | \psi_{k,a} \ket
\bra \psi_{k,a} | \rho | \psi_{k,a} \ket
\bra \psi_{k,a} | y \ket \\
& = & \sum_{u,v = 1}^d
\bra x | \psi_{k,a} \ket
\bra \psi_{k,a} | u \ket
\bra u | \rho | v \ket
\bra v | \psi_{k,a} \ket
\bra \psi_{k,a} | y \ket \nonumber
\ee
Furthermore
\be
\bra x | \psi_{k,a} \ket =
\bra x | G^{k} H^{a} | \theta \ket =
{1 \over \sqrt{d}} \exp \bigg[{2 \pi i k x \over d}\bigg]
\,
\exp \bigg[{2 \pi i a x^2 \over 2d^{2}}\bigg]
\ee
Substituting into (\ref{(a,k)}) gives
\be\label{eqn2}
& &  \bra x | E_{k,a} \rho E_{k,a} | y \ket \\
&  & = {1 \over d^2} \sum_{u,v =1}^d
\bra u | \rho | v \ket \,
\exp \bigg[{2 \pi i  k \over d} (x + v - y - u) \bigg]
\,
\exp \bigg[{2 \pi i  a \over 2d^{2}} (x^2 + v^2 - y^2 -
u^2) \bigg] \nonumber
\ee
When the right side of (\ref{eqn2}) is substituted in (\ref{eqn1}),
the sum over $k$ gives zero unless $x+v-y-u$ is an integer
multiple of $d$. Since $x,v,y,u$ vary between $1$ and $d$,
the only possible values are 
\be\label{sum-k}
x + v - y - u = 0, \,  d, \, -d
\ee
Similarly, the sum over $a$ gives zero unless
$x^2 + v^2 - y^2 - u^2$ is an integer multiple of
$2 d^2$. In this case the only possibility is
\be\label{sum-a}
x^2 + v^2 - y^2 - u^2 = 0
\ee
Consider first the case that (\ref{sum-k}) gives
\be\label{sum-k-d}
x + v - y - u =  d
\ee
Let $\gamma = y + u$, then $x + v = \gamma + d$,
and hence
\be
2 \leq \gamma \leq d
\ee
Elementary bounds then lead to
\be
x^2 + v^2 > \gamma^2 > y^2 + u^2
\ee
which shows that there can be no simultaneous solution
of (\ref{sum-a}) and (\ref{sum-k-d}). A similar argument holds for the case
\be
x + v - y - u =  - d
\ee
The remaining case in (\ref{sum-k}) can be written as
\be\label{case3}
x - u =  y - v
\ee
and also (\ref{sum-a}) can be written as
\be\label{case4}
(x - u) (x + u) = (y - v) (y + v)
\ee
It follows that the only simultaneous solutions
of the equations (\ref{case3}) and (\ref{case4}) are
$x=y, u=v$ and $x=u, y=v$. Hence if $x \neq y$ the left side of
(\ref{eqn1}) gives $\bra x | \rho | y \ket$, while if
$x = y$ the sum gives $\sum_{u} \bra u | \rho | u \ket =
\Tr \rho = 1$.
\QED

\medskip

Combining Lemma \ref{lemma5} and Lemma \ref{lemma6} we
arrive at the convex decomposition
of $\Delta_{\lambda}$, which expresses the depolarizing channel in terms of 
the phase-damping channels
$\Phi_{\lambda}^{(a)}$:
\be\label{bigdecomp}
\Delta_{\lambda}(\rho) & = & 
{\lambda \over 1 + (d-1) \lambda}
\,\, {1 \over 2 d} \,\, \sum_{a=1}^{2 d^2} \Phi_{\lambda}^{(a)}(\rho) \\
& + & {1 - \lambda \over 1 + (d-1) \lambda} \,\, {1 \over 2 d^3}\,\,
\sum_{k=1}^d \sum_{a=1}^{2 d^2} 
\big( G^{*} \big)^{k} \Phi_{\lambda}^{(a)}(\rho) G^{k} \nonumber
\ee
Furthermore each state $| \psi_{k,a} \ket$
defined in (\ref{def:psi}) is uniform, and hence the phase-damping
channels $\Phi_{\lambda}^{(a)}$ defined in (\ref{def:newPhi}) are
also uniform. This completes the proof of Lemma \ref{lemma2}.

\subsection{$d = 2$: the qubit depolarizing channel}
It is useful to look in detail at the familiar case $d=2$.
A general state $\rho$ can be written as a $2 \times 2$ hermitian
matrix
\be\label{2drho}
\rho = \pmatrix{a & c \cr
\overline{c} & b \cr}
\ee
The depolarizing channel (\ref{def:Delta}) acts by
\be\label{depol2}
\Delta_{\lambda}(\rho) = \pmatrix{
{\lambda}_{+} a + {\lambda}_{-} b & \lambda c \cr
\lambda \overline{c} & {\lambda}_{-} a + {\lambda}_{+} b  \cr},
\ee
where we have defined
\be\label{def:lam+-}
{\lambda}_{\pm} = {1 \pm \lambda \over 2}
\ee
Also the `intermediate' channel $\Omega_{\lambda}$ defined in
(\ref{def:Omega}) acts by
\be\label{omega2}
\Omega_{\lambda}(\rho) = \pmatrix{
{\lambda}_{+} a + {\lambda}_{-} b & {\lambda}_{+} c \cr
{\lambda}_{+} \overline{c} & {\lambda}_{-} a + {\lambda}_{+} b  \cr}
\ee
The first diagonal unitary matrix $G$ defined in (\ref{def:G}) is just
\be
G = \pmatrix{-1 & 0 \cr
0 & 1 \cr} 
= - \sigma_z
\ee
So the first relation (\ref{convex1}) becomes
\be
\Delta_{\lambda} = {2 \lambda \over 1 + \lambda} \Omega_{\lambda}
+ {1 - \lambda \over 1 + \lambda} \, {1 \over 2} \,
\bigg[ \Omega_{\lambda} + \sigma_z \Omega_{\lambda} \sigma_z \bigg]
\ee
which can be easily verified using (\ref{depol2}) and (\ref{omega2}).

The second diagonal unitary matrix $H$ defined in (\ref{def:H})
is now
\be
H = \pmatrix{\exp \Big[{\pi i / 4}\Big] & 0 \cr
0 & -1 \cr} 
\ee
There are eight phase-damping channels defined in (\ref{def:newPhi}).
Four of these can be written in terms of the usual Pauli matrices:
\be
{\Phi}_{\lambda}^{(2)}(\rho) = {\Phi}_{\lambda}^{(6)}(\rho) =
{\lambda}_{+} \rho + {\lambda}_{-} \sigma_{y} \rho \sigma_{y}
\ee
\be
{\Phi}_{\lambda}^{(4)}(\rho) = {\Phi}_{\lambda}^{(8)}(\rho) =
{\lambda}_{+} \rho + {\lambda}_{-} \sigma_{x} \rho \sigma_{x}
\ee
The others can be written in terms of the following
Pauli-type matrices:
\be\label{def:tau}
\tau = \pmatrix{0 & \exp \Big[{\pi i / 4}\Big] \cr
\exp \Big[- {\pi i / 4}\Big]  & 0 \cr} , \quad \quad
\overline{\tau} = \pmatrix{0 & \exp \Big[- {\pi i / 4}\Big] \cr
\exp \Big[{\pi i / 4}\Big]  & 0 \cr} 
\ee
The relations are
\be
{\Phi}_{\lambda}^{(1)}(\rho) = {\Phi}_{\lambda}^{(5)}(\rho) =
{\lambda}_{+} \rho + {\lambda}_{-} \tau \rho \tau
\ee
\be
{\Phi}_{\lambda}^{(3)}(\rho) = {\Phi}_{\lambda}^{(7)}(\rho) =
{\lambda}_{+} \rho + {\lambda}_{-} \overline{\tau} \rho \overline{\tau}
\ee
The second convex decomposition (\ref{convex2}) now reads
\be
\Omega_{\lambda} = {1 \over 8} \sum_{a=1}^{8}
{\Phi}_{\lambda}^{(a)}
\ee

There is a lot of redundancy in
the final decomposition (\ref{bigdecomp}), which now has 24 terms on the right side.
In fact $\Omega_{\lambda}$ can be written as a convex combination of
just two uniform phase-damping channels, namely
\be
\Omega_{\lambda} = {1 \over 2} \Big[
{\Phi}_{\lambda}^{(2)} + {\Phi}_{\lambda}^{(4)} \Big],
\ee
and this allows $\Delta_{\lambda}$ to be written as a convex
combination of just four phase-damping channels.
There may be a similar redundancy in (\ref{bigdecomp}) for $d > 2$.

\section{The phase-damping channel}
In this section we will establish Lemma \ref{lemma3} for
the product channel $\Phi_{\lambda} \otimes I$.
Without loss of generality we will choose the basis 
${\cal B} = \{ | i \ket \}$, so that 
$\Phi_{\lambda}$ acts on a state by simply scaling all off-diagonal
entries by the same factor $\lambda$,
as in (\ref{diag-basis}):
\be
\Phi_{\lambda}(\rho) = \lambda \rho +
(1 - \lambda) \sum_{i=1}^d | i \ket \bra i | \, \bra i | \rho | i \ket
\ee

The product channel $\Phi_{\lambda} \otimes I$
acts on bipartite states $\rho_{12}$ defined
on ${\bf C}^d \otimes {\bf C}^{d'}$ for some dimension $d'$.
It will  be convenient to view these states as $d \times d$ block matrices,
where each
block is itself a $d' \times d'$ matrix.
Furthermore there is a convenient factorization of these blocks,
which can be derived by the following observation.
Let us write $\sqrt{\rho_{12}} = (V_{1} \dots V_d)$ where
each $V_i$ is a   $d d' \times d'$ matrix. Then we have
\be\label{factor1}
\rho_{12} & = & \Big( \sqrt{\rho_{12}}\Big)^{*} \, \sqrt{\rho_{12}} \\ \nonumber
 & = & \pmatrix{V_1^{*} V_1 & \dots & V_{1}^{*} V_d \cr
\vdots & \ddots & \vdots \cr
V_{d}^{*} V_1 & \dots & V_{d}^{*} V_d \cr} 
\ee
Recall the definition of $\rho_{2}^{(i)}$ in (\ref{def:rho^i}).
With our choice of basis here, the matrix $E_i \otimes I$ is simply
the orthogonal projector onto the $i^{\rm th}$ block on the
main diagonal, hence
\be\label{rho-i}
\rho_{2}^{(i)} = V_{i}^{*} \, V_{i}
\ee

The key to deriving the bound (\ref{bound1}) is to rewrite the
factorization (\ref{factor1}) as follows:
\be\label{factor2}
\rho_{12} & = & \pmatrix{V_{1}^{*} & 0 & \dots & 0 \cr
0 & V_{2}^{*} & \dots & 0 \cr
\vdots &  & \ddots & \vdots \cr
0 &  & \dots & V_{d}^{*} }\,
M \,
\pmatrix{V_{1} & 0 & \dots & 0 \cr
0 & V_{2} & \dots & 0 \cr
\vdots &  & \ddots & \vdots \cr
0 &  & \dots & V_{d} }
\ee
where $M$ is the $d \times d$ block matrix
\be
M =\pmatrix{I' & \dots & I' \cr
\vdots & \ddots & \vdots \cr
I' & \dots & I' }
\ee
and $I'$
is the $d d' \times d d'$ identity matrix.
Recall the state $| \theta \ket$ defined in (\ref{def:theta}).
Using this we can rewrite $M$ as the product state
\be
M = d \, \Big(| \theta \ket \bra \theta | \Big) \, \otimes I'
\ee
Furthermore the simple action of the phase-damping
channel $\Phi_{\lambda}$ implies that it acts on (\ref{factor2})
in the following way:
\be
(\Phi_{\lambda} \otimes I) (\rho_{12}) & = &
\pmatrix{V_1^{*} V_1 & \dots & \lambda \, V_{1}^{*} V_d \cr
\vdots & \ddots & \vdots \cr
\lambda \, V_{d}^{*} V_1 & \dots & V_{d}^{*} V_d} \\
&  & \cr \nonumber
& = & \pmatrix{V_{1}^{*} & 0 & \dots & 0 \cr
0 & V_{2}^{*} & \dots & 0 \cr
\vdots &  & \ddots & \vdots \cr
0 &  & \dots & V_{d}^{*} }\,
(\Phi_{\lambda} \otimes I) (M) \,
\pmatrix{V_{1} & 0 & \dots & 0 \cr
0 & V_{2} & \dots & 0 \cr
\vdots &  & \ddots & \vdots \cr
0 &  & \dots & V_{d} } \nonumber
\ee
Let us define
\be
A = \pmatrix{V_{1}V_{1}^{*} & 0 & \dots & 0 \cr
0 & V_{2}V_{2}^{*} & \dots & 0 \cr
\vdots &  & \ddots & \vdots \cr
0 &  & \dots & V_{d} V_{d}^{*}}
\ee
and
\be
B = (\Phi_{\lambda} \otimes I) (M) =
d \, \Big( \Phi_{\lambda}(| \theta \ket \bra \theta |) \Big)
\, \otimes I'
\ee
Then  $(\Phi_{\lambda} \otimes I) (\rho_{12})$ has the same spectrum as
the matrix $A^{1/2} B A^{1/2}$. Therefore
\be\label{bound4-1}
\Tr \bigg( (\Phi_{\lambda} \otimes I) (\rho_{12}) \bigg)^p =
\Tr \bigg( A^{1/2} B A^{1/2} \bigg)^p
\ee
Now we use the Lieb-Thirring inequality \cite{LT}, which states that for all
$p \geq 1$
\be\label{L-T}
\Tr \bigg( A^{1/2} B A^{1/2} \bigg)^p \leq
\Tr \bigg( A^{p/2} B^p A^{p/2} \bigg)
= \Tr \bigg( A^p B^p \bigg)
\ee
The matrix $ A^{p}$ is block diagonal:
\be\label{A-p}
A^{p} = \pmatrix{(V_{1}V_{1}^{*})^p & 0 & \dots & 0 \cr
0 & (V_{2}V_{2}^{*})^p  & \dots & 0 \cr
\vdots &  & \ddots & \vdots \cr
0 &  & \dots & (V_{d} V_{d}^{*})^p}
\ee
Furthermore
\be\label{B-p}
B^p = d^p \, \Big( \Phi_{\lambda}(| \theta \ket \bra \theta |) \Big)^p\,
\otimes I'
\ee
Explicit calculation shows that the diagonal entries of
$d^p \, \Big( \Phi_{\lambda}(| \theta \ket \bra \theta |) \Big)^p$
are all
equal to $(1 - \lambda)^p +
\Big[ (d \lambda + 1 - \lambda)^p - (1 - \lambda)^p \Big]/d$.
Comparing this with (\ref{nu4dep}), and substituting (\ref{A-p}) and
(\ref{B-p}) into the right side of (\ref{L-T}) we get
\be\label{evalL-T}
\Tr \Big( A^{p} B^p \Big) =
d^{(p-1)}\, \bigg( \nu_{p}(\Delta_{\lambda}) \bigg)^p \,
\sum_{i=1}^d \Tr (V_{i} V_{i}^{*})^p
\ee
Now recall (\ref{rho-i}), and also notice that
for all $i = 1, \dots, d$
\be\label{iden5}
\Tr (V_{i} V_{i}^{*})^p  = \Tr (V_{i}^{*} V_{i})^p
= \Tr \bigg( \rho_{2}^{(i)} \bigg)^p
\ee
Combining (\ref{L-T}), (\ref{evalL-T})  and
(\ref{iden5}) gives the bound (\ref{bound1}).
\QED

\section{The additivity of $\chi^{*}$}
The proof of Theorem \ref{thm2} uses the representation of
$\chi^{*}$ as a min-max of relative entropy, combined with an
entropy bound derived from Lemma \ref{lemma3}. The relative
entropy  representation
was derived by  Ohya, Petz and Watanabe \cite{OPW}
and Schumacher and Westmoreland \cite{SW2}.  Recall that the
relative entropy of two states $\rho$ and $\omega$ is defined as
\be
S(\rho, \omega) = \Tr \rho ( \log \rho - \log \omega)
\ee
The OPWSW representation for the Holevo capacity of the
channel $\Psi$ is
\be\label{OPWSW}
\chi^{*}(\Psi) & = & \inf_{\omega} \sup_{\rho}
S \Big(\Psi(\rho), \Psi(\omega) \Big) \\
& = &  \sup_{\rho}
S \Big(\Psi(\rho), \Psi(\omega^{*}) \Big)
\ee
where the state $\omega^{*}$ that achieves the infimum in (\ref{OPWSW})
is the optimal average input state from the channel. For the depolarizing
channel this optimal average is $(1/d) \, I$, that is
the totally mixed state.  For the product channel
$\Delta_{\lambda} \otimes \Psi$,
the Holevo quantity $\chi^{*}(\Delta_{\lambda} \otimes
\Psi)$ can be upper bounded by choosing
$(1/d) \, I \otimes \omega^{*}$ as the average input
state. This leads to the following inequalities:
\be
\chi^{*}(\Delta_{\lambda}) + \chi^{*}(\Psi)
\leq \chi^{*}(\Delta_{\lambda} \otimes \Psi) \leq
\sup_{\tau_{12}}
S \Big((\Delta_{\lambda} \otimes \Psi)(\tau_{12}),
(1/d) \, I \otimes \Psi(\omega^{*}) \Big)
\ee
In order to prove additivity we will combine this with the following
result.

\medskip
\begin{lemma}\label{lemma7}
For all bipartite states $\tau_{12}$,
\be\label{bound2}
S \Big( (\Delta_{\lambda} \otimes \Psi)(\tau_{12}),
(1/d) \, I \otimes \Psi(\omega^{*}) \Big) \leq
\chi^{*}(\Delta_{\lambda}) + \chi^{*}(\Psi)
\ee
\end{lemma}

\medskip
\noindent{\bf Proof:}
\par\noindent
The left side of (\ref{bound2}) can be rewritten as
\be\label{expand1}
S \Big( (\Delta_{\lambda} \otimes \Psi)(\tau_{12}),
(1/d) \, I \otimes \Psi(\omega^{*}) \Big)
& =&
- S \Big((\Delta_{\lambda} \otimes \Psi)(\tau_{12}) \Big) \\
& + &
\log d - \Tr \Psi(\tau_2) \log \Psi(\omega^{*}) \nonumber
\ee
where $\tau_2$ is the reduced density matrix of $\tau_{12}$.
From here on we follow the steps in the proof of
Theorem \ref{thm3}.
First, by Lemma \ref{lemma1}  we can assume without loss of generality that
$\tau_{1} = {\Tr}_{2} (\tau_{12})$ is diagonal.
Second, notice that the channel $\Delta_{\lambda}$ appears on the right
side of (\ref{expand1}) only in the first term. Therefore
Lemma \ref{lemma2} and concavity of the entropy
imply that it is sufficient to establish the bound
\be\label{bound2-1}
S \Big( (\Phi_{\lambda} \otimes \Psi)(\tau_{12}),
(1/d) \, I \otimes \Psi(\omega^{*}) \Big) \leq
\chi^{*}(\Delta_{\lambda}) + \chi^{*}(\Psi)
\ee
where $\Phi_{\lambda}$ is a uniform phase-damping channel and
where $\tau_1$ is diagonal.

Next we apply (\ref{bound1}) with $\rho_{12} = (I \otimes \Psi)
(\tau_{12})$, and take the derivative at $p=1$ to get
\be\label{bound3}
S \Big((\Delta_{\lambda} \otimes \Psi)(\tau_{12}) \Big)
& \geq & S_{\rm min}(\Delta_{\lambda}) - \log d \\
& - & \sum_{i=1}^d x_i \log x_i + \sum_{i=1}^d
x_i S \bigg(\Psi \Big({1 \over x_i} \tau_{2}^{(i)} \Big) \bigg) \nonumber
\ee
where $x_i = \Tr \Big(\tau_{2}^{(i)}\Big)$, and as usual
\be\label{recall1}
\tau_{2}^{(i)} = {\Tr}_{1} \Big[ (E_i \otimes I) \tau_{12} \Big]
\ee
Since $\tau_1$ is diagonal and $\Phi_{\lambda}$ is uniform, it
follows that
\be
x_i = {1 \over d}
\ee
for all $i = 1, \dots, d$, hence
$\sum x_i \log x_i = - \log d$. Also recall the evaluation of
$\chi^{*}(\Delta_{\lambda})$ in (\ref{chi4dep}).
Hence (\ref{bound3}) can be written as
\be\label{bound3-1}
S \Big((\Delta_{\lambda} \otimes \Psi)(\tau_{12}) \Big)
 \geq  - \chi^{*}(\Delta_{\lambda})  + \log d + {1 \over d} \, \sum_{i=1}^d
S \bigg(\Psi \Big(d \, \tau_{2}^{(i)} \Big) \bigg)
\ee

Furthermore, since the
projections $E_i$ in (\ref{recall1}) constitute an orthonormal
basis it follows that
\be\label{sum1}
\sum_{i=1}^d \tau_{2}^{(i)} = {\Tr}_{1} \Big[ (I \otimes I) \tau_{12} \Big]
= \tau_2
\ee
Therefore the left side of (\ref{bound2-1}) can be rewritten as
in (\ref{expand1}) to get
\be\label{expand2}
S \Big( (\Phi_{\lambda} \otimes \Psi)(\tau_{12}),
(1/d) \, I \otimes \Psi(\omega^{*}) \Big) & = &
- S\Big((\Phi_{\lambda} \otimes \Psi) (\tau_{12}) \Big) \\
& + &
\log d - {1 \over d} \,
\sum_{i=1}^d \Tr \Psi(d \tau_{2}^{(i)}) \log \Psi(\omega^{*}) \nonumber
\ee

Combining (\ref{expand2}) with (\ref{bound3-1})
we get
\be\label{bound4}
S \Big( (\Delta_{\lambda} \otimes \Psi)(\rho_{12}),
(1/d) \, I \otimes \Psi(\omega^{*}) \Big)
& \leq & \chi^{*}(\Delta_{\lambda}) \\
& + & {1 \over d} \, \sum_{i=1}^d  S \Big(
\Psi(d \, \tau_{2}^{(i)}), \Psi(\omega^{*}) \Big) \nonumber
\ee
Recall that $\Tr (d \tau_{2}^{(i)}) = 1$. Therefore it follows
from (\ref{OPWSW}) that for each $i =1, \dots, d$
\be
S \Big(
\Psi(d \, \tau_{2}^{(i)}), \Psi(\omega^{*}) \Big) \leq
\chi^{*}(\Psi)
\ee
and hence (\ref{bound4}) implies (\ref{bound2-1}). \QED

\section{Conclusions and discussion}
We have  presented the proof of a long-conjectured property of the
$d$-dimensional depolarizing channel
$\Delta_{\lambda}$, namely that its capacity for
transmission of classical information can be achieved with
product signal states and product measurements. This result
follows as a consequence of  several additivity results which
we prove for the product channel $\Delta_{\lambda} \otimes \Psi$
where $\Psi$ is an arbitrary channel.
The principal result is the proof of the AHW conjecture for
the matrix $p$-norm, for all $p \geq 1$,
from which we deduce the additivity of minimal entropy and of the Holevo
quantity. The argument presented here is
a generalization of the method used earlier by the author
to prove similar
results for all unital qubit channels, and involves
re-writing the depolarizing channel
as a convex combination of other simpler channels, which we
refer to as phase-damping channels.

If the additivity conjecture for the Holevo quantity
is true for all channels,  then there must be a general
argument which can be used to provide a proof, and presumably this would
give a different method of proof for Theorem \ref{thm1}.
However it is known that the AHW conjecture is not true
in general \cite{Werner}, and indeed
it is an interesting problem to determine the class of
channels for which it does hold.
As a consequence, it may be that the method of this paper gives the
most direct route to the proof of the AHW property for the
depolarizing channel.
It is expected that the same method can be applied to prove the
AHW result for
a class of $d$-dimensional channels, and this question is under study.

\bigskip
{\bf Acknowledgements}
The author is grateful to Wayne Sullivan and John Lewis for
valuable discussions leading to the bound in
Lemma \ref{lemma3}. This work was partially supported by 
National Science Foundation Grant DMS--0101205, and by
Science Foundation
Ireland under the National Development Plan.

{~~}

\end{document}